# Quantitative comparisons to promote inquiry in the introductory physics lab

N.G. Holmes[1] & D.A. Bonn[2]

[1]Physics Department, Stanford University, Stanford, CA
[2]Department of Physics & Astronomy, University of British Columbia, Vancouver, BC

In a recent report, the American Association of Physics Teachers has developed an updated set of recommendations for curriculum of undergraduate physics labs.[1] This document focuses on six major themes: constructing knowledge, modeling, designing experiments, developing technical and practical laboratory skills, analyzing and visualizing data, and communicating physics. These themes all tie together as a set of practical skills in scientific measurement, analysis, and experimentation. In addition to teaching students *how* to use these skills, it is important for students to know *when* to use them so that they can use them autonomously. This requires, especially in the case of analytical skills, high-levels of inquiry behaviours to reflect on data and iterate measurements, which students rarely do in lab experiments.[2,3] Often, they perform lab experiments in a plug-and-chug frame, procedurally completing each activity with little to no sensemaking.[2,3] An emphasis on obtaining true theoretical values or agreement on individual measurements also reinforces inauthentic behaviours such as retroactively inflating measurement uncertainties.[2] This paper aims to offer a relatively simple pedagogical framework for engaging students authentically in experimentation and inquiry in physics labs.

A primary hurdle to students' inauthentic inquiry in labs is their conceptual understanding of measurement and uncertainty. These concepts have been framed around two contrasting paradigms: set and point paradigms.[4] Buffler and colleagues define these terms as follows: "The point paradigm is characterized by the notion that each measurement could in principle be the true value. ... The set paradigm is characterized by the notion that each measurement is only an approximation to the true value and that the deviation from the true value is random" (p. 1139).[4] The point paradigm emphasizes the importance of any single piece of data, assigning special importance to the individual measured value. In contrast, the set paradigm emphasizes only the importance of a collection of data, recognizing that an individual measured value is only an estimate of the physical quantity being measured.

In an introductory lab, this manifests itself in a variety of ways. First, the use of the term "error" creates a pedagogical problem when teaching a novice experimenter. Students may interpret measurement errors as actual mistakes that have caused the measured value to differ from a "true value,"[5] which falls under the point paradigm. This notion extends into a belief that measurement errors could be reduced to zero and a perfect measurement of the true value could be made (presumably, by scientists in a lab).[5,6] Thus, point-like thinking encourages students to interpret errors (synonymously with uncertainties) as measures of accuracy, rather than precision. In fact, many students in our course have indicated that they had previously compared measurements to true or actual values through the equation,

$$\%Error = \frac{|Actual - Measured|}{Actual} \cdot 100\%. \qquad (1)$$

The use of the word error here indeed refers to a quantitative measure of accuracy, as they take the difference between their measured value and the actual or true value as a percent fraction of the actual or true value. It is clear, then, why students may subsequently misinterpret measurement errors (meaning uncertainty) as literal errors (meaning mistakes), expressing the deviation of a measurement from its theoretical value.[6] This issue further aggravates the separation since it completely ignores the measurement uncertainty in the comparison values.

Another common method for comparing measurements is through overlapping uncertainty ranges. This dichotomous comparison (either the ranges overlap indicating agreement or do not overlap indicating disagreement) reinforces point-like thinking in another way as it ignores the continuous probability distribution that characterizes each measurement (that is, beyond one unit of uncertainty). This issue is reinforced by the graphical representation of error bars, which provide a finite range around the measured value, inadvertently implying the full distribution around the measurement. The mathematical notation itself (using the ± symbol) is guilty of reinforcing this idea. It also supports more extreme versions of point-like thinking, especially if the language (plus or minus) is interpreted literally, as in the solution to the quadratic equation.

**The *t'*-score framework**

With these misconceptions about measurement and uncertainty and our goals for engaging students in meaningful reflection in the lab, we aimed to move comparisons to a continuous rather than dichotomous scale. The continuous scale is defined by how many units of standard uncertainty the two measurement values differ (for example, 1σ, 2σ, or 3σ differences). This scale is defined by:

$$t' = \frac{A - B}{\sqrt{\delta_A^2 + \delta_B^2}}, \qquad (2)$$

where A and B are two separate measurements (or means of two sets of measurements) and $\delta_A$ and $\delta_B$ are their uncertainties, respectively. The *t'*-score takes the difference between the two measurements and normalizes by the combined uncertainty of the difference.[8] As such, the *t'*-score gives a quantitative measure of how different the two measurements are relative to their uncertainties, similar to the 'sigma' levels used in particle physics, for example. Structurally, this index is similar to *Student's t* statistic and measures of effect size, but we do not make inferences from the statistic on the *t* distribution. Indeed, if the measurements were sample means from populations with the same variance, the *t'*-score would be equivalent to *Student's t* for comparing independent samples.

In addition to putting comparisons on a continuous and set-like scale, rather than a dichotomous and point-like scale, the *t'*-score provides a quantitative way to evaluate the quality of one's data and to suggest follow up experiments or measurements (Table 1). A *t'*-score around 1 could suggest that the two measurements are indeed similar (A is close to B) or that measurement uncertainties are very large. As such, small *t'*-scores encourage improved follow-up measurements aimed to reduce uncertainties. A large *t'*-score around 3 could mean students have underestimated their uncertainties (which is unlikely, in our experience), or that the two measurements are very different, that they have made an error, or that there are limitations or unjustified approximations to the physical model at play. This engages students in the challenging reflection task of identifying possible systematic errors or evaluating the physical models. Regardless of the size of the *t'*-score, students have a structure for conducting follow-up measurements and iterating to improve their results.

| *t'*-score | Interpretation of measurements | Follow-up investigation |
|---|---|---|
| $|t'|<1$ | Unlikely different, uncertainty may be overestimated | Improve measurements, reduce uncertainty |
| $1<|t'|<3$ | Unclear whether different | Improve measurements, reduce uncertainty |
| $3<|t'|$ | Likely different | Improve measurements, correct systematic errors, evaluate model limitations or approximations |

Table 1. Interpretations of and follow-up behaviours from a *t'-score* comparison between two measurements in the SQILab.

This full inquiry framework (comparing, reflecting, iterating to improve) is part of the Structured Quantitative Inquiry Labs (SQILabs). Although students conduct a fairly prescribed experiment initially, they are never told how to improve their measurement in subsequent iterations, providing a constrained experimental design opportunity. This autonomous design phase provides a more authentic scientific inquiry experience, developing students' scientific abilities.[9] It also offers individualized Socratic guidance[10] without the instructor needing to intervene, making the inquiry process self-driven. We will provide an example of the *t'*-score framework in use during an early pendulum experiment to show evidence of its impact.

**Pendulum for Pros Experiment**

The Pendulum for Pros experiment takes place in the second week of our introductory college physics course, immediately following instruction on standard deviation[11], standard uncertainty in the mean, and *t'*-scores. To put these skills to use, students are asked to measure and compare the period of a pendulum through timing measurements at two different amplitudes (angles of 10° and 20°). The physical equation dictating this process usually presented to students in class is,

$$T = 2\pi\sqrt{L/g} \qquad (3)$$

where *T* is the period, *L* is the length of the pendulum, and *g* is the acceleration due to gravity. This equation suggests that the period is independent of the amplitude of the swing and thus the measurements at 10° and 20° should not differ. The derivation of this equation, however, makes an approximation that assumes *sinθ≈θ*, which is only valid for small angles. If students are familiar with the approximation in the formula, they are often unsure as to where this equation is valid in terms of measurement (what constitutes a small angle?). With reasonable precision (uncertainties around 0.1% of the measured values, obtained through trials of approximately 50-swing measurements), the two measurements at 10° and 20° are distinct, with the 20° angle being 'large enough' to vary from this approximation.

In the SQILab, students were instructed to measure the period of a pendulum at 10° and 20°, compare the two values using a *t'*-score, reflect on and interpret the meaning of the *t'*-score, and then conduct follow-up measurements based on the *t'*-score comparison. While they were told to make measurements at two prescribed angles, they were not told *how* to make the measurements nor what follow-up measurements to make. In order to avoid comparisons to 'true' or 'theoretical' values, students were not given equation 3, nor were they asked to calculate a predicted value through measurements of the length of the pendulum.

As an example of how the process works, we will demonstrate one pair of students' progress through the experiment. Table 2 shows their three measurement attempts. First they made 10 single-period trials and calculated means and standard uncertainties in the means for the two angle measurements. Comparing these using a *t'*-score gave them a value of 0.11. Based on the *t'*-score framework, it is unclear whether the measurements are different, so the students should improve their measurement to reduce their uncertainties. Their design was to let the pendulum swing 10 times between measurements (so they only start and stop the stopwatch once every 10 swings rather than once per swing), thus decreasing their uncertainty by a factor of 10. This gave an increased *t'*-score of 2.35, suggesting they should further improve the measurement to confirm whether they differ. In their third attempt, they increased the number of swings per measurement to 20, thus reducing their uncertainty in the previous measurement by a factor of 2, and increasing their *t'*-score to 3.66, at which point they concluded that the values were different.

| Attempt | Students' design | T at 10° (s) | T at 20° (s) | |*t'*|-score |
|---|---|---|---|---|
| 1 | Measure single period 10 times | 1.83 ±0.08 | 1.81 ±0.08 | 0.18 |
| 2 | Measure 10 periods 5 times | 1.823±0.008 | 1.850±0.008 | 2.39 |
| 3 | Measure 20 periods 5 times | 1.830±0.004 | 1.851±0.004 | 3.71 |

Table 2. Sample data sets produced by a group of students in the Pendulum experiment. The three iterations represent their progression through the activity as they attempt to improve their measurement quality.

A year before the *t'*-score framework was introduced in our course, a similar version of this experiment was used with a focus on making comparisons between measured data points using overlapping uncertainty ranges. In this session, students compared the periods of the pendulum at three different angles (5°, 10°, and 25°). Without the SQILab, nearly all students made a single trial of ten-swing measurements of the pendulum, achieving an average uncertainty of 0.017±0.002s. By the end of the session in the SQILab, students' measurements involved, on average, three trials of 40-swing measurements, achieving an average uncertainty of 0.0035±0.0005s. Table 3 summarizes these distinctions.

| Cohort | Number of trials | Number of swings | Average uncertainty (s) |
|---|---|---|---|
| Non-SQILab | 1.008±0.008 | 10±0 | 0.017±0.002 |
| SQILab | 3.0±0.2 | 40±2 | 0.0035±0.0005 |

Table 3. Summary of students' behaviours each year during similar experiments comparing the period of the pendulum at different amplitudes. Data represent mean values within the class samples (Non-SQILab, *n*=121; SQILab, *n*=90) and standard uncertainties in the means.

It is clear, then, that the explicit instructions to iterate and improve measurements, in an autonomous way, improved the quality of students' data. How students interpret and evaluate these results is also important. Figure 1 is an image of the final conclusion from one of the students in the sample SQILab group. After their third comparison, the student identifies (confesses) that this result is the opposite of what he had expected. He expected, from class, that the pendulum periods would be the same. He does, however, recognize the limitations of the model, namely that it is only valid for small angles. He then begins to tie together the physical model, the mathematical approximation, and the reality of experimental measurement as he discusses how 20° is not a 'small enough' angle when precision is high.

Figure 1. One student's conclusion at the end of the Pendulum for Pros experiment discusses the conflict between the outcome they expected and the results they have obtained.

Through this experience, the students see that they can make precise measurements that are better than the approximations they see in class. This is a non-trivial experience given that many students expect to make poor quality measurements, especially poorer than expert physicsts.[2,6,12] The students also explore the limitations of physical models, recognizing that the physical world is often more complicated than what is presented. This experience sets them up for future model-based discussions about approximations that are made, either about the physical models or the measurement system.[13] This also suggests to students, somewhat implicitly, that they should attempt high quality measurements in the lab, since the results may not be as expected.

With the $t'$-score framework presented to students early in the year, each experiment should involve opportunities for comparisons, some of which involve common systematic errors to be corrected through iteration, limitations or approximations of physical models to be evaluated, or appropriate agreement through high quality measurements. With explicit instructions to compare, reflect, and iterate through the $t'$-score framework early in the year, these behaviours become habit and routine during the experiments as the explicit instructions are faded throughout the course. In addition, opportunities to evaluate models expand as the $t'$-score framework gets mapped on to weighted least-squares fitting for continuous data, allowing this structure to be used across the lab curriculum.

In summary, we find this framework provides a straightforward structure for encouraging authentic scientific inquiry behaviours in the lab, shifting students to more expert-like epistemological frames and improving their understanding of measurement, uncertainty, and models.

**Acknowledgements**

The authors would like to acknowledge the helpful comments from referees. This research was supported by the UBC Carl Wieman Science Education Initiative.